\documentclass[preprint,prd,aps,showpacs,showkeys,nofootinbib]{revtex4-1}
\usepackage{amssymb}
\usepackage{amsmath}
\usepackage{graphicx}
\usepackage{subfigure}
\usepackage{float}

\begin{document}


\title{Inflection point inflation and dark energy in supergravity}

\author{Tie-Jun Gao}
\email{gaotj@itp.ac.cn}
\affiliation{State Key Laboratory of Theoretical Physics, Institute of
Theoretical Physics, Chinese Academy of Sciences, P.O. Box 2735,
Beijing 100190, China}

\author{Zong-Kuan Guo}
\email{guozk@itp.ac.cn}
\affiliation{State Key Laboratory of Theoretical Physics, Institute of
Theoretical Physics, Chinese Academy of Sciences, P.O. Box 2735,
Beijing 100190, China}

\begin{abstract}
We consider an inflection point inflationary model in supergravity with a single chiral superfield and show that the predicted values of the scalar spectral index and tensor-to-scalar ratio are consistent with the Planck 2015 results. In this model supersymmetry is strongly broken after inflation, which results in a non-SUSY de-Sitter vacuum responsible for the recent accelerated expansion of the Universe.
\end{abstract}

\keywords{} \pacs{}

\maketitle

\section{Introduction\label{sec1}}

Cosmological inflation is now getting established by all precise observational data such as the WMAP~\cite{ref1} and Planck space missions~\cite{ref0}. The full-mission Planck observations of temperature and polarization anisotropies of the cosmic microwave background radiation constrain the spectral index of curvature perturbations and the tensor-to-scalar ratio to be $n_s=0.9655\pm0.0062$ and $r_{0.002}<0.10$ at $95\%$ confidence level~\cite{ref0}, respectively, which are consistent with the analysis of Planck 2013~\cite{ref2}.

However, the nature of inflation remains an open question in cosmology. An interesting framework for inflation model building is to embed the inflationary models into a more fundamental theory of quantum gravity, and it is natural to consider supergravity. Some inflationary models have been constructed in supergravity~\cite{ref11,ref12,ref13}, most of which however suffer from the so-called $\eta$ problem~\cite{ref14}. The F-term of the potential is proportional to $e^{|\Phi|^2}$, which gives a contribution of the slow-roll parameter $\eta$ and breaks the slow-roll condition. Several methods are proposed to solve this problem~\cite{ref141,ref142,ref143,ref144}.
One way to overcome such obstacles was to add an extra chiral superfield $S$ and to use a shift-symmetric K\"{a}hler potential $K(\Phi+\bar{\Phi},\bar{S}S)$ \cite{ref15,ref16,ref17,ref23,ref24}. During inflation, the superfield $S$ is stabilized
at $S = 0$. In these models there are two superfields, with four scalar degrees of freedom, but only one of them to be the inflaton field while the others never participate in the cosmological evolution.

Recently, in Refs.~\cite{ref19,ref20}, Ketov and Terada propose a new class of inflationary models with only one  chiral superfield $\Phi$. Generally, the superfield $\Phi$ is decomposed into a real part $\phi$ and an imaginary one $\chi$, such as
\begin{eqnarray}
&&\Phi=\frac{1}{\sqrt{2}}(\phi+i\chi).
\end{eqnarray}
Following~\cite{ref19}, we consider the following logarithmic K\"{a}hler potential
\begin{eqnarray}
&&K=-3\ln\left[1+\frac{\Phi+\overline{\Phi}+\zeta(\Phi+\overline{\Phi})^4}{\sqrt{3}}\right].
\label{kp}
\end{eqnarray}
Since it is invariant under the shift $\Phi\rightarrow\Phi+i C$ with a real parameter $C$, the imaginary component  $\chi$ does not appear in the  K\"{a}hler potential, which could play the role of the inflaton field. The quartic term  serves to stabilize the field $\chi$ during  the main part of inflation at $\phi\approx0$. As shown in Appendix C of Ref. \cite{ref19}, although the quadratic and cubic terms in the Kahler potential (2) are allowed by the symmetries, the coefficients of such terms can be suppressed by tuning the coupling between the superfield $\Phi$ and other superfields. In this paper, we shall take $\zeta=1$ for simplicity.

A complete cosmological model must include the stages of both early acceleration and later acceleration of the Universe,
so in the framework of supergravity with the K\"{a}hler potential~\eqref{kp}, the authors of Ref.~\cite{ref21} consider a linear superpotential with a small constant and a quadratic superpotential with a linear correction, respectively. At the end of inflation, the field will roll to a non-SUSY AdS vacuum. By a small modification of the  parameters in the theory, one can uplift the vacuum to non-SUSY dS vacuum with a tiny cosmological constant $V_0\sim10^{-120}$, which does not violate the no-go theorem~\cite{ref22}.

In the framework of MSSM, a successful inflection point inflation is for the first time realized in the gauge invariant flat directions udd or LLe \cite{ref23}. In such a model, the fine tuning and reheating are discussed in detail in Ref. \cite{ref24}. A solution of the fine tuning problem was proposed in a minimal extension of MSSM in \cite{ref25}. Due to an attractor behavior towards the inflection point, the initial condition for the MSSM inflation can be naturally realized \cite{ref26}. Recently it is pointed out that inflection point inflation can yield large tensor-to-scalar ratios \cite{ref27,ref28}. The purpose of the present paper is to investigate a class of supergravity models motivated by superstring compactification and supersymmetric particle phenomenology beyond the Standard Model \cite{ref19,ref20,ref21}. The inflaton may belong to a hidden sector and can decay into the SM particles after inflation \cite{ref29}.

In this paper, we shall consider the possibility to construct an inflection point inflationary model in supergravity with a single chiral superfield. We shall focus on a superpotential of the form $W=m(\Phi^3+ae^{i\theta}\Phi+be^{i\rho})$. We study the inflaton dynamics and show that the predicted  scalar spectral index and tensor-to-scalar ratio can lie within the $1\sigma$ confidence region allowed by the results of Planck 2015.  After the end of inflation, the potential has a global non-SUSY minimum, as found in Ref.~\cite{ref21}, one can uplift the potential to have the desirable dS vacuum with $V_0\sim10^{-120}$ by fine-tuning the model parameters.

The rest of this paper is organized as follows. In the next section, we setup the inflection point inflationary model in supergravity. In Section 3, we investigate the inflaton dynamics of the model. In Section 4 we study the vacuum structure of the model and explore the parameter space to give the desirable inflation and dark energy. The last section is devoted to summary.

\section{Setup of the inflection point inflation \label{sec2}}

In the supergravity theory with the K\"{a}hler potential~\eqref{kp}, for an arbitrary choice of the superpotential, the kinetic term of the field $\Phi$ is given by~\cite{ref21}
\begin{eqnarray}
&&L_{kin}=\frac{3(1-24\sqrt{3}\zeta\phi^2-8\sqrt{2}\zeta\phi^3+32\zeta^2\phi^6)}{(\sqrt{3}+\sqrt{2}\phi+4\zeta\phi^4)^2}\partial_\mu\Phi\partial^\mu\bar{\Phi}.
\end{eqnarray}
The coefficient of the kinetic term does not depend on $\chi$ and it is positive definite when $\phi\in(-0.159,0.152)$ for $\zeta=1$. So $\phi$ is confined in a narrow interval, and $\chi$ plays the role of the inflaton field.

The potential is determined by a given superpotential  $W$ as well as  K\"{a}hler potential, which is given by
\begin{eqnarray}
&&V=e^K\Big[D_iW(K^{-1})^{i\bar{j}}(D_jW)^*-3|W|^2\Big],
\label{infp}
\end{eqnarray}
where
\begin{eqnarray}
&&D_iW=\partial_iW+(\partial_iK)W,
\end{eqnarray}
and $(K^{-1})^{i\bar{j}}$ is the inverse of the K\"{a}hler metric
\begin{eqnarray}
&&K^{i\bar{j}}=\frac{\partial^2K}{\partial\Phi_i\partial\bar{\Phi}_{\bar{j}}}.
\end{eqnarray}

In order to give  an inflection point inflation as well as a tiny cosmological constant after inflation, we consider the superpotential of the form
\begin{eqnarray}
&&W=m(\Phi^3+ae^{i\theta}\Phi+be^{i\rho}),
\label{sp}
\end{eqnarray}
where  the  coefficient $m$ is real, $a$ and $b$  are positive without loss of generality, $\theta$ and $\rho$  are the phase of the coefficients. In order to study further inflation and vacuum structure after inflation in the parameter space, let us first consider the case of $\rho=0$ for simplicity. Later we will show that the value of $\rho$ cannot affect the inflationary predictions.

\begin{figure}\small
  \centering
   \includegraphics[width=4in]{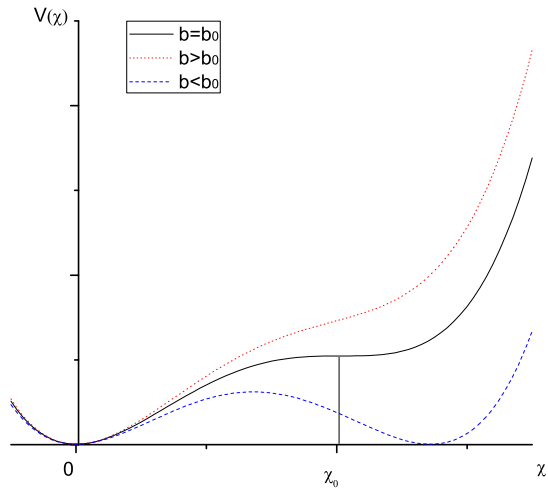}
     \caption{The inflation potential $V(\chi)$ when $\phi$ is stabilized at zero.}
    \label{fig-sp}
\end{figure}

Substituting the superpotential~\eqref{sp} and K\"{a}hler potential~\eqref{kp} into~\eqref{infp}, one can get the potential. Because of the shift symmetry, there is no imaginary component $\chi$ in the K\"{a}hler potential, so the potential is considerably flat along the $\chi$ direction and thus $\chi$ becomes an inflaton candidate. As shown above, the real component $\phi$ is stabilized at zero during inflation. Therefore, we can  set $\phi=0$ and  obtain  the scalar potential of $\chi$
\begin{eqnarray}
&&V(\chi)=m^2\left(\frac{9}{4}{\chi ^4} - \sqrt 6 a\sin \theta {\chi ^3} + 3(\sqrt 3 b - a\cos \theta ){\chi ^2} + {a^2} - 2\sqrt 3 ab\cos \theta\right).
\label{pchi}
\end{eqnarray}
The cubic term leads to a negative contribution when $a\sin \theta>0$. The inflation potential is shown in Fig.~\ref{fig-sp}.

If the parameters satisfy the relation
\begin{eqnarray}
&&b=\frac{a \left(9 \cos \theta+2 a \sin ^2\theta  \right)}{9 \sqrt{3}},
\end{eqnarray}
there are two  minima at $\chi=0$ and at $\chi=\frac{{4a\sin \theta }}{{3\sqrt 6 }}$, respectively, as shown in Fig.~\ref{fig-sp}. As $b$ increases, the minimum at $\chi=\frac{{4a\sin \theta }}{{3\sqrt 6 }}$ is uplifted. In this case, for a large initial value of $\chi$, the inflaton field may be trapped in the false vacuum.

An interest case is that if the parameters satisfy the relation
\begin{eqnarray}
&&b=b_0=\frac{{a\left( {4\cos \theta  + a\sin^2 {\theta }} \right)}}{{4\sqrt 3 }},
\label{brelation}
\end{eqnarray}
the minimum of the potential at $\chi=\chi_0= a\sin \theta /{\sqrt 6 }$ becomes equal to the local maximum, and thus the false vacuum disappears. This point is the so-called inflection point. At this point, the inflation potential is
\begin{eqnarray}
V(\chi_0) &&=\frac{1}{2}{a^2}{m^2}\left( {2 - 4\cos^2 {\theta} - a\cos \theta\sin^2 {\theta} + \frac{1}{{24}}{a^2}\sin^4 {\theta}} \right).
\end{eqnarray}
Both the first and second derivatives of $V$ vanish at $\chi_0$. We will see shortly that since there is a flat plateau around the inflection point, the predicted spectral index of curvature perturbations as well as the tensor-to-scalar ratio can lie within the $1\sigma$ confidence region allowed by Planck 2015.
When $a\sin \theta\rightarrow0$ and then $\chi_0\rightarrow0$, the chaotic inflationary model is reproduced.

In addition, it is known that large Hubble-induced mass corrections to the inflation potential can ruin the flatness of the potential. However, such corrections may not be a serious issue in the context of inflection point inflation. The point is that certain relations among the parameters are needed to be satisfied in order to find an inflection point in the potential. Adding such terms will make these relations different, but one may still be able to realize inflection point inflation in Ref. \cite{ref30}.

In this paper we focus on the inflation potential with inflection point. Since the parameters satisfy the relation~\eqref{brelation}, there are only three free parameters $a, m$ and $\theta$. The inflation potential~\eqref{pchi} becomes
\begin{eqnarray}
V(\chi)= \frac{m^2 }{4}\left(9 \chi ^4-4\sqrt{6} a  \sin \theta \chi ^3+3 a^2 \sin ^2\theta \chi ^2 -2 a^2 \left(a \sin ^2\theta  \cos \theta +2 \cos 2 \theta \right) \right).
\end{eqnarray}

\section{Slow-roll inflation   \label{sec3}}

The slow-roll parameters are defined as
\begin{eqnarray}
&&\epsilon \equiv \frac{1}{2}\left(\frac{V'(\chi)}{V(\chi)}\right)^2 ,\nonumber\\
&& \eta \equiv \frac{V''(\chi)}{V(\chi)}.
\end{eqnarray}
To first order in the slow-roll approximation, the scalar spectral index and tensor-to-scalar ratio are given by
\begin{eqnarray}
&&n_s \simeq 1-6\epsilon+2\eta ,\nonumber\\
&&r \simeq 16\epsilon.
\end{eqnarray}
The $e$-folding number during inflation is given by
\begin{eqnarray}
&&N=\int^{\chi_i}_{\chi_f}\frac{V}{V'}d\chi ,
\end{eqnarray}
where the field value at the end of inflation $\chi_f$ is determined by Max$\{\epsilon(\chi_f),\eta(\chi_f)\}=1$.

The parameter $m$ is constrained by the amplitude of curvature perturbations
\begin{eqnarray}
&&\Delta_R^2=\frac{V}{24\pi^2\epsilon}.
\end{eqnarray}
Using the maximum likelihood value $\Delta^2_R(k_0)=2.19\times10^{-9}$ from the Planck 2015 data and set $\theta=1.55$ and $a=96$ we can get $m\sim 5.76\times10^{-8}$.

\begin{figure}\small
  \centering
   \includegraphics[width=4in]{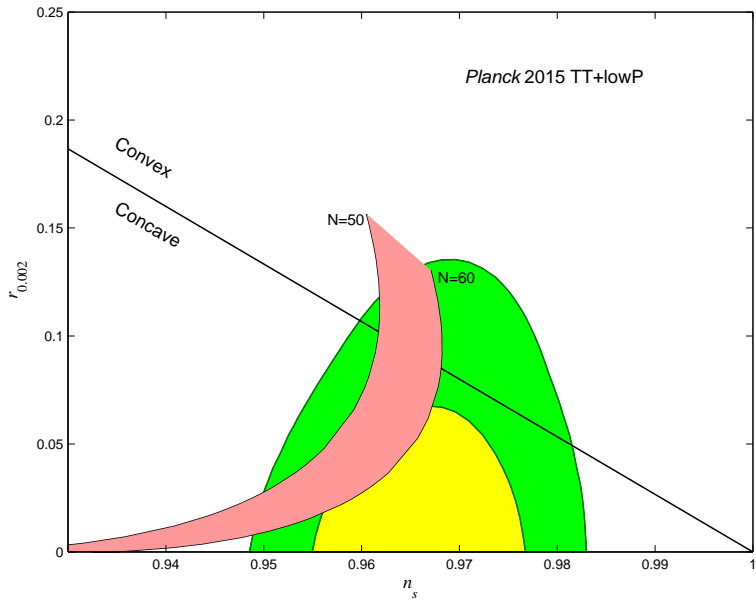}
     \caption{(color online)The $n_s-r$ region (pink region) predicted by the model with the $e$-folding number from $N = 50$ (left boundary line) to $N = 60$
(right boundary line). The contours are the marginalized joint 68\% and 95\% confidence level regions for $n_s$ and $r$ at the pivot scale $k_*= 0.002$ Mpc$^{-1}$ from the Planck 2015 TT+lowP data.}
    \label{fig-nr}
\end{figure}

In order to give an appropriate vacuum structure, the parameters are strongly restricted (see Fig.~\ref{fig-at}), so only one parameter is free. For example, if one set $\theta=1.55$, then one can get the desired dS vacuum only if $a\sim96$. In this case the model predicts that $n_s=0.968$ and $r=0.081$ for the $e$-folding number of $N=60$. Fig.~\ref{fig-nr} shows the $n_s-r$ region (pink region) predicted by the model with the $e$-folding number from $N=50$ (left boundary line) to $N=60$ (right boundary line). The contours are the marginalized joint 68\% and 95\% confidence level regions for $n_s$ and $r$ at the pivot scale $k_*=0.002$ Mpc$^{-1}$ from the Planck 2015 TT+lowP data.  It can be seen that the predictions are consistent with the Planck 2015 results.

\section{Vacuum structure of the potential   \label{sec4}}

\begin{figure}
\begin{minipage}[t]{0.5\linewidth}
\centering
\includegraphics[width=2.7in]{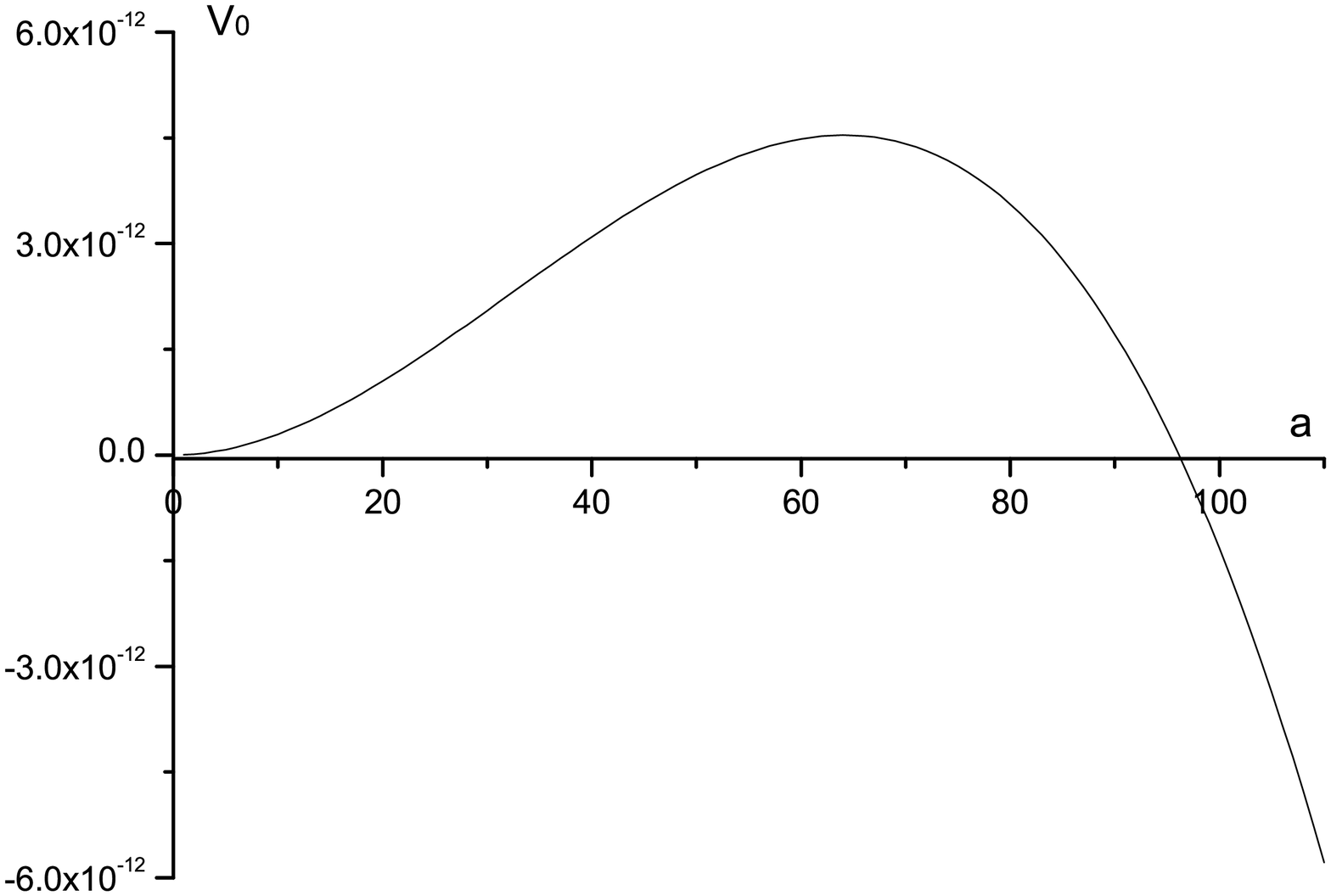}
\end{minipage}%
\begin{minipage}[t]{0.5\linewidth}
\centering
\includegraphics[width=2.7in]{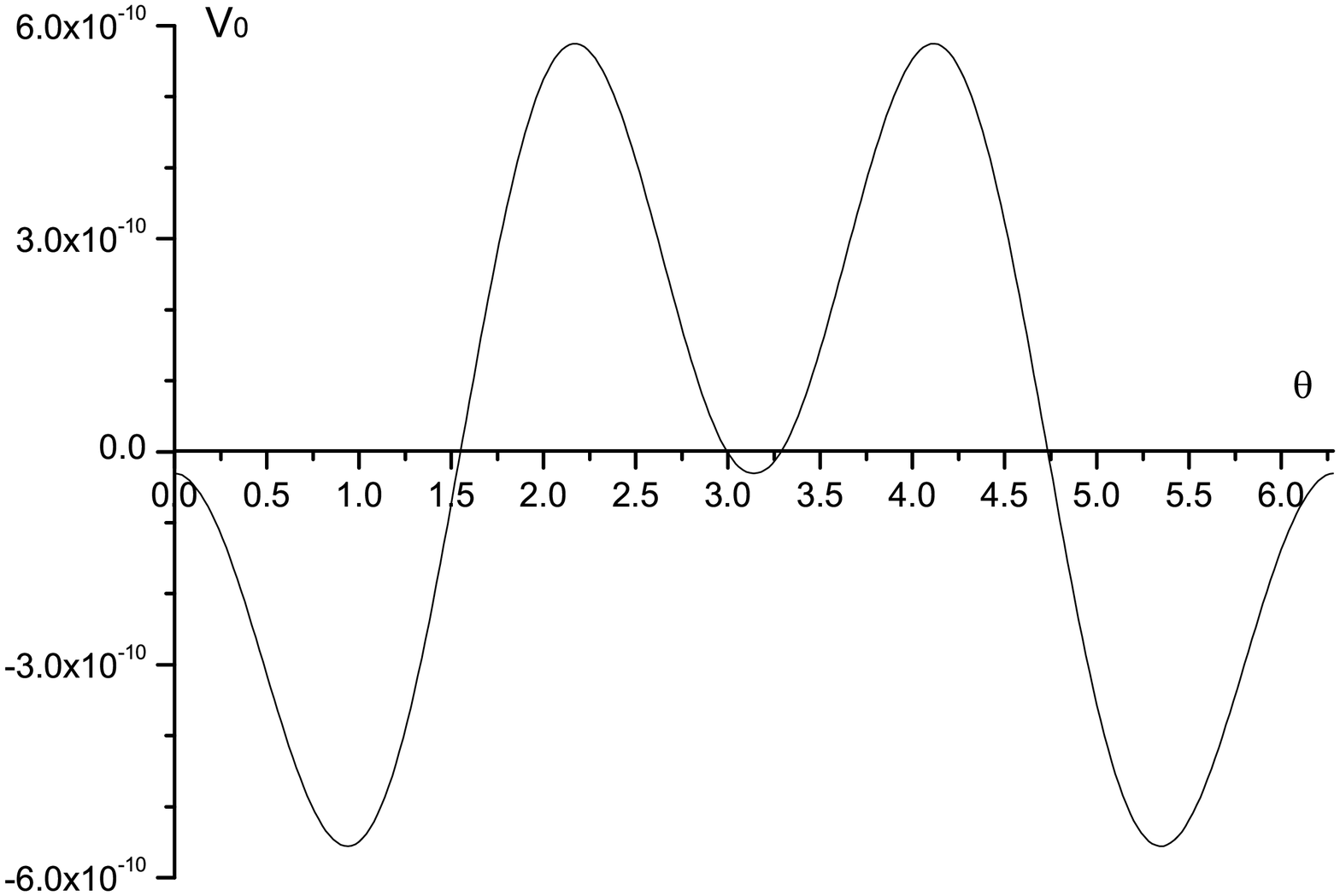}
\end{minipage}
\caption{The value of the cosmological constant in the minimum as a
function of the parameter $a$ for $\theta=1.55$ (left panel) and as a
function of $\theta$ for $a=96$ (right panel). We choose $\zeta=1, m=5.76\times10^{-8}$.}
\label{fig-v0at}
\end{figure}

 After inflation the field $\chi$ rolls towards $\chi = 0$. However, the global minimum of the potential is no longer at $\chi=0$ and $\phi=0$, but  the location shifts a little in the $\phi$ direction. Such a small deviation from $\phi=0$ cannot affect the inflationary predictions. One can change the values of $a$ and $\theta$ to uplift a non-SUSY AdS vacuum to a non-SUSY dS vacuum, which does not violate the no-go theorem. For example, for $a\approx96$ and $\theta\approx1.55$, there is a global non-SUSY minimum at $\chi=0$ and $\phi\approx3.4\times10^{-5}$. The desired dS vacuum with $V_0\sim10^{-120}$ can be uplifted by a minuscule change of the parameters $a$ and $\theta$. Although it requires a fine tuning, it's not a major problem in the landscape scenario of string theory.
 Fig.~\ref{fig-v0at} shows the value of the cosmological constant in the minimum as a function of the parameter $a$ for $\theta=1.55$ (left panel) and as a function of $\theta$ for $a=96$ (right panel).

\begin{figure}\small
  \centering
   \includegraphics[width=4in]{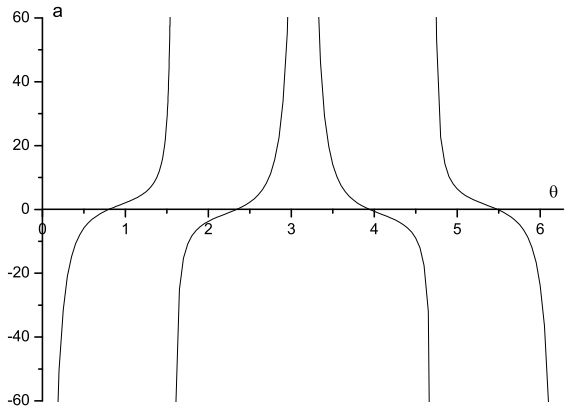}
     \caption{The parameter space of $a$ and $\theta$ that gives a non-SUSY Minkowski minimum for $\zeta=1$.}
    \label{fig-at}
\end{figure}

We can see that as the parameters $\theta$  increases or $a$ decreases, it can give rise to a transition between  AdS and  dS vacuum, passing through Minkowski vacuum. Therefore, in order to get an appropriate vacuum structure, we can fine tune $a$ for a given $\theta$. The relation between $\theta$ and $a$ which can give a non-SUSY  Minkowski vacuum are shown in Fig.~\ref{fig-at}. Then the inflationary predictions of $n_s$ and $r$ depend only on the value of $\theta$.

In addition, the supersymmetry is strongly broken in the minimum of the potential. For $\zeta=1,\theta=1.55$, the superpotential at the minimum is $W \sim 7.6867\times10^{-5}$ and the gravitino mass is $m_{3/2} \sim 7.6864\times10^{-5}$, in Planck units, i.e. $m_{3/2} \sim 1.87051\times10^{14}$ GeV, which is one order of magnitude higher then in Ref.~\cite{ref21}. Such a scale is much higher than the usual predictions of the supersymmetry breaking sclae in supergravity phenomenology.

\begin{figure}\small
  \centering
   \includegraphics[width=4in]{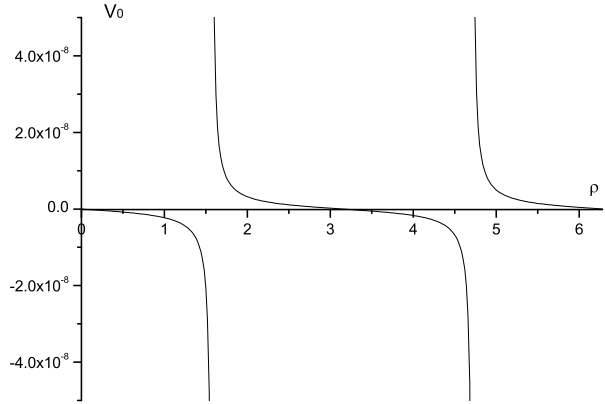}
     \caption{The value of the cosmological constant in the minimum as a
function of the parameter $\rho$ for $\theta=1.55$ and  $a=96$. We choose $\zeta=1, m=5.76\times10^{-8}$.}
    \label{fig-v0r}
\end{figure}

When the parameter $\rho$ in superpotential is changed, the desired dS vacuum can be obtained by changing $a$ and $\theta$. Moreover, we have checked that the inflationary predictions are independent of $\rho$.
Fig.~\ref{fig-v0r} show the value of the cosmological constant as a function of $\rho$ for $\theta=1.55$ and $a=96$.

\section{Summary \label{sec5}}

The complete cosmological model including both the early acceleration and the present acceleration of the Universe has been investigated in the framework of supergravity with a single chiral superfield.
In this model, the inflection point inflation in the $\chi$ direction has successfully been constructed using the logarithmic K\"{a}hler potential~\eqref{kp} and the cubic superpotential~\eqref{sp}. The inflationary predictions of the model are consistent with the Planck 2015 results. Such predictions in the $n_s-r$ plane do not overlap those of hilltop quartic inflation and have small overlap with those of natural inflation. Future measurements of temperature and polarization anisotropies of the cosmic microwave background radiation can test and distinguish them.

After inflation, the non-SUSY minimum of the potential can be uplifted to a non-SUSY dS vacuum with vanishingly small vacuum energy $V_0\sim10^{-120}$ without violating the no-go theorem by fine tuning the model parameters.
In this model, supersymmetry after inflation is strongly broken and the predicted value of the gravitino mass is  much higher than the often assumed TeV mass range.

\begin{acknowledgments}
This work was supported in part by the National Natural Science Foundation of China No.11175225 and No.11335012.
\end{acknowledgments}

\end{document}